\documentclass[final,5p,times,twocolumn]{elsarticle}
\usepackage{amsmath,amsfonts,amssymb,array}
\usepackage{bm,bbm,bbm,bbold}
\usepackage{enumerate}
\usepackage{dsfont}
\usepackage{float}
\usepackage{graphicx}
\usepackage{longtable}
\usepackage{multirow}
\usepackage{slashed,subfigure}
\usepackage{xcolor}
\usepackage[colorlinks = true,linkcolor = blue]{hyperref} 
\definecolor{darkgreen}{rgb}{0.0, 0.5, 0.0}

\def\pTVeto{p_{T}^{\rm Veto}}
\def\varVeto{\varepsilon(p_T^{\rm Veto})}

\def\ab{{\rm ~ab}}
\def\fb{{\rm ~fb}}
\def\pb{{\rm ~pb}}
\newcommand{\invfb}{{\rm ~fb^{-1}}}
\newcommand{\invab}{{\rm ~ab^{-1}}}

\def\GeV{{\rm ~GeV}}
\def\TeV{{\rm ~TeV}}
\def\met{{\rm MET}}

\journal{Physics Letters B}

\hypersetup{bookmarks=true,unicode=true,pdftoolbar=true,pdfmenubar=true,
  pdffitwindow=false,pdfstartview={FitH},
  pdftitle={Pascoli, Ruiz, Weiland - Safe Jet Vetoes - Phys. Lett. B786, 106 (2018) [arXiv:1805.09335]},pdfauthor={Pascoli, Ruiz, Weiland},
  pdfsubject={Subject},pdfcreator={Creator},pdfproducer={Producer},
  pdfkeywords={Beyond the Standard Model}{Jet Veto}{Collider Physics}{Neutrino Physics},
  pdfnewwindow=true,colorlinks=true
}

\begin{document}

\begin{frontmatter}

\title{Safe Jet Vetoes}

\author[IPPP]{Silvia Pascoli}  
\ead{silvia.pascoli@durham.ac.uk}

\author[IPPP]{Richard Ruiz}  
\ead{richard.ruiz@durham.ac.uk}

\author[IPPP]{Cedric Weiland}  
\ead{cedric.weiland@durham.ac.uk}

\address[IPPP]{Institute for Particle Physics Phenomenology, Department of Physics, Durham University, Durham DH1 3LE, U.K.}


\begin{abstract}
Central jet vetoes are powerful tools for reducing QCD background in measurements and searches for electroweak and colorless, new physics processes in hadron collisions.
In this letter, we report the key findings of a new philosophy to designing searches for such phenomena at hadron colliders,
one designed and centered around a dynamical jet veto instead a static veto applied independently of other selection criteria.
Specifically, we investigate the theoretical and phenomenological consequences of setting the jet veto scale to the transverse momentum $(p_T)$
of the leading charged lepton $\ell$ in multi-lepton processes on an event-by-event basis.
We consider the case of a TeV-scale heavy neutrino $N$ decaying to the trilepton final state and find the following: 
(i) Perturbative uncertainties associated with the veto greatly reduce due to tying the veto scale to the hard process scale.
(ii) The signal efficiency for passing the veto jumps to $\gtrsim95\%$ and exhibits little-to-no dependence on the neutrino mass scale.
(iii) Top quark and ``fake'' lepton rejection capabilities also improve compared to only vetoing heavy flavor-tagged jets above a fixed $p_T$.
This results in an increased sensitivity to active-sterile neutrino mixing by approximately an order of magnitude  over the LHC's lifetime.
For a Dirac neutrino with mass $m_N = 150-1000$ GeV and the representative active-sterile mixing hypothesis $\vert V_{e4}\vert = \vert V_{\tau 4}\vert$ with $\vert V_{\mu 4}\vert=0$,
we find that LHC experiments 
can probe $\vert V_{e4}\vert^2, \vert V_{\tau 4}\vert^2 \lesssim 6\times10^{-4} - 8\times10^{-3}$, 
surpassing the global upper limit for $m_N < 450$ GeV, with $\mathcal{L}=3$ ab$^{-1}$ of data at $\sqrt{s}=14$ TeV.
Due to the color structures of the heavy $N$ production mechanisms considered, we argue that our results hold broadly for other color-singlet processes.
\end{abstract}

\begin{keyword}
Beyond the Stardard Model \sep Jet Veto \sep Collider Physics \sep Neutrino Physics \sep arXiv:1805.09335
\end{keyword}

\end{frontmatter}

\section{Introduction}\label{sec:Intro}

Jet vetoes, i.e., the rejection of events with jets above a transverse momentum threshold $(p_T^{\rm Veto})$,
are incredibly powerful tools for reducing QCD backgrounds in measurements and searches for 
electroweak (EW) and colorless new physics processes at hadron colliders.
In conjunction with heavy quark flavor-tagging, 
jet vetoes are among the most widely used techniques by experiments at the Large Hadron Collider (LHC).

Theoretically, however, jet vetoes are, simply put, complicated.
Foremost, select arguments of the Collinear Factorization Theorem~\cite{Collins:1989gx,Stewart:2009yx,Collins:2011zzd,Becher:2012qa}, 
i.e., the master equation for computing hadronic scattering rates, do not hold in the presence of a veto due to its exclusive nature.
More precisely, jet vetoes receive corrections from Glauber exchanges, e.g., double parton scattering/multiple parton interactions,
which are beyond the theorem's formal accuracy nor are presently known to factorize;
for further details, see Refs.~\cite{Dasgupta:2007wa,Stewart:2009yx,Collins:2011zzd,Becher:2012qa,Gaunt:2014ska,Zeng:2015iba}.
In addition, for color-singlet processes occurring at a mass scale $Q$,
vetoes give rise to logarithmic dependencies on $\pTVeto$ of the form $\alpha_s(\pTVeto)\log(Q^2/p_T^{{\rm Veto}\, 2})$.
While usually perturbative in practice, such contributions and uncertainties  are sufficiently large that high-accuracy resummation, 
either analytically~\cite{Banfi:2012jm,Becher:2013xia,Stewart:2013faa,Meade:2014fca,Jaiswal:2014yba,Becher:2014aya} 
or by parton showers means~\cite{Monni:2014zra}, is necessary to reproduce EW data.
Moreover, the effectiveness of vetoes in searches for new high-mass particles is considerably hindered 
by the higher predisposition of higher mass objects to generate QCD radiation than lighter objects~\cite{Tackmann:2016jyb,Fuks:2017vtl}.

In this letter, we report on a particular jet veto implementation, which we describe as a ``safe jet veto,'' 
that addresses the latter two concerns.
Specifically, for final states with multiple charged leptons, $pp\to n\ell+X$, $\ell\in\{e,\mu,\tau\}$, 
we set on an event-by-event basis the value of $\pTVeto$ to be the $p_T$ of the leading (highest $p_T$) charged lepton.
Dynamical jet vetoes, such as the one we propose, have been considered previously 
in the context of EW boson production~\cite{Denner:2009gj,Nhung:2013jta,Frye:2015rba}, but only for computational convenience.
Here, we demonstrate that they can be successfully used in a much broader class of experimental searches,
including searches for new, high-mass colorless particles as well as events with $\tau$ leptons decaying hadronically.
We find impressive improvement over traditional, fixed-$p_T$ jet vetoes.

We report three key findings: 
(i) Perturbative QCD uncertainties associated with the veto greatly reduce due to tying the veto scale to the hard process scale, 
i.e., by effectively converting a two-scale problem into a one-scale problem.
(ii) The signal efficiency for passing the dynamical veto is very high and exhibits little-to-no dependence on the signal mass scale, 
unlike with static vetoes, where efficiency drops with increasing mass scale.
(iii) Top quark and ``fake'' lepton rejection capabilities also improve compared to only vetoing heavy flavor-tagged jets.

\begin{figure}[!t]
\includegraphics[width=0.45\textwidth]{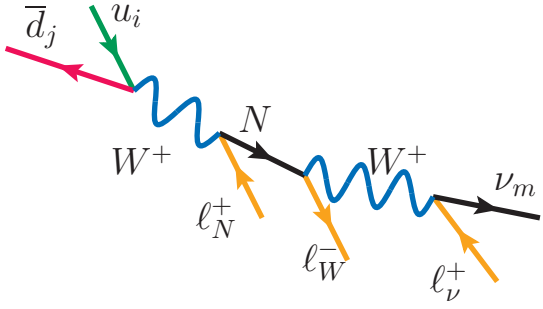}
\caption{
Born-level diagram of a heavy Dirac neutrino $N$ produced via the DY process with a subsequent decay to the trilepton final-state.
Drawn with \texttt{JaxoDraw}~\cite{Binosi:2003yf}.
}
\label{fig:feynman_DYX_Nl_3lX}
\end{figure}

To illustrate these results, we have investigated the production in proton collisions of a hypothetical, 
heavy colorless particle, namely a heavy neutrino $(N)$, that decays to a trilepton final state.
We consider heavy neutrino production via both the charged current Drell-Yan~\cite{Keung:1983uu} process 
$pp\to \ell_N N \to \ell_N \ell_W W \to \ell_N \ell_W \ell_\nu \nu$,
as shown in Fig.~\ref{fig:feynman_DYX_Nl_3lX}, as well as the $(W\gamma)$ vector boson fusion process~\cite{Dev:2013wba,Alva:2014gxa,Degrande:2016aje}.
Here and below, the subscript $X = N,W,\nu$ on the charged lepton $\ell_X$ denotes the particle produced in association with $\ell_X$.
Due to the color structures of the production mechanisms considered, this case study is broadly representative of many new physics scenarios.

This letter continues in the following manner:
We first briefly summarize the relevant ingredients of our heavy neutrino model in Sec.~\ref{sec:theory} and computational inputs in Sec.~\ref{sec:mc}.
We define and discuss our signal processes in Sec.~\ref{sec:signalProc}.
In Sec.~\ref{sec:jetVeto}, we discuss how the proposed veto scheme impacts differently the signal and backgrounds processes, which leads to our findings (i)-(iii).
The impact of the veto on searches for heavy neutrinos at the LHC, as well as a brief outlook, are then presented in Sec.~\ref{sec:results}.
Finally, we conclude in Sec.~\ref{sec:conclusions}.
For a more extensive collection of results, we refer readers to Ref.~\cite{PRW18:toAppear}.

\section{Simplified Heavy Neutrino Model}\label{sec:theory}

TeV-scale neutrino mass models are well-motivated and typically assume the existence of singlet massive fermions, which mix with the active ones. 
At colliders such models commonly predict new colorless resonances that decay readily to multiple charged 
leptons~\cite{Keung:1983uu,delAguila:2007qnc,Atre:2009rg,Deppisch:2015qwa,Cai:2017mow}.
Hence, to demonstrate the power of dynamical jet vetoes and how they improve searches for new particles, 
we consider a simplified model that extends the SM by a single colorless fermion, a heavy neutrino.
In this simplified $(3+1)$ model with Dirac neutrinos, and working in a basis where the charged lepton mass and flavor eigenstates coincide,
neutrino flavor eigenstates $(\nu_\ell)$ 
are related to light $(\nu_m)$ and heavy $(N)$ mass eigenstates by~\cite{Atre:2009rg} 
\begin{equation}
 \nu_{\ell} = \sum_{m=1}^{3} U_{\ell m}\nu_{m} + V_{\ell 4} N.
\end{equation}
The above should be understood as applying to the left-handed components of spinors.
For TeV-scale heavy $N$, global fits to precision EW precision and low-energy observables, such as tests of lepton universality and CKM unitarity, 
constrain $\vert V_{\ell 4}\vert \lesssim 0.021-0.075$~\cite{Antusch:2014woa,Fernandez-Martinez:2016lgt} at $2\sigma$.
After EWSB and in the mixed basis to first order in $V_{\ell 4}$, the relevant couplings of $N$ to SM fields are given by
\footnote{This form of the mixing and of the interaction Lagrangian, in particular  
 the Higgs coupling, is inspired by low-scale seesaw models.}
\begin{eqnarray}
  \mathcal{L}_{\rm Int.} = 
  &-& \frac{g}{\sqrt{2}}W^+_\mu		\sum_{\ell=e}^\tau 		~\overline{N} ~V_{\ell 4}^* 	~\gamma^\mu P_L\ell^-\nonumber\\
    &-& \frac{g}{2\cos\theta_W}Z_\mu	\sum_{\ell=e}^\tau 		~\overline{N} ~V_{\ell 4}^*	~\gamma^\mu P_L\nu_\ell\nonumber\\
  &-& \frac{g m_N}{2M_W} h	\sum_{\ell=e}^\tau 		~\overline{N} ~V_{\ell 4}^*	 P_L\nu_\ell + \text{H.c.}\,,
  \label{eq:Lagrangian}
\end{eqnarray}
with $g$ being the usual $\mathrm{SU}(2)_L$ coupling constant.
For the impact of jet vetoes on the larger particle spectrum of a full neutrino mass model, see Ref.~\cite{PRW18:toAppear}.

\section{Computational Setup}\label{sec:mc}

To compute our signal and background processes, we use a Dirac neutrino variant of the 
NLO in QCD-accurate~\cite{Hahn:2000kx,Degrande:2014vpa}
\texttt{HeavyNnlo}~\cite{Alva:2014gxa,Degrande:2016aje} FeynRules~\cite{Alloul:2013bka,Christensen:2008py,Degrande:2011ua} model file.
LO(+PS) and NLO+(PS) event generation is performed by \texttt{MadGraph5\_aMC@NLO} v2.6.2$\beta$~\cite{Alwall:2014hca},
with parton shower matching (including QED radiation) via \texttt{Pythia8} v230~\cite{Sjostrand:2014zea}
using the CUETP8M1 ``Monash*'' tune~\cite{Skands:2014pea}, 
and particle-level reconstruction to standard~\cite{Alwall:2006yp} Les Houches Event files by \texttt{MadAnalysis5} v1.6.33~\cite{Conte:2012fm}.
Jets are clustered according to the anti-$k_T$ algorithm~\cite{Cacciari:2008gp} with $R=1$ (unless otherwise specified), 
as implemented in \texttt{FastJet}~v3.2.0~\cite{Cacciari:2011ma}.
Computations at these accuracies are matched to the \texttt{NNPDF 3.1 NLO+LUXqed} PDF set~\cite{Bertone:2017bme} 
due to its LUXqed-based $\gamma$-PDF~\cite{Manohar:2016nzj,Manohar:2017eqh}.
As argued in Refs.~\cite{Martin:2004dh,Martin:2014nqa,Alva:2014gxa,Degrande:2016aje},
such a formalism provides the most appropriate description of the  $W^{*}\gamma\to N\ell^\pm$ fusion process.
NLO+NNLL(veto) rates are calculated within the framework of Soft-Collinear Effective Field Theory (SCET)~\cite{Bauer:2000yr,Bauer:2001yt,Beneke:2002ph},
using Ref.~\cite{Alwall:2014hca,Becher:2014aya};
the computation is matched with the \texttt{NNPDF 3.1 NNLO+LUXqed} PDF set 
to avoid double counting of $\mathcal{O}(\alpha_s^2)$ contributions in the NNLL resummation.
Approximate NNNLL(threshold) rates use the SCET-based calculation of Ref.~\cite{Ruiz:2017yyf}, following Refs.~\cite{Becher:2006nr,Becher:2006mr},
and the \texttt{NNPDF 3.1 NNLO+LUXqed} PDF set.

Where reported, we show the dependence on the factorization and renormalization scales, as well as the hard and soft scales if applicable,
by varying the default scales discretely by $0.5\times$ and $2.0\times$.
We evolve PDFs and $\alpha_s(\mu)$ using the LHAPDF v6.1.6~\cite{Buckley:2014ana}.

\section{Heavy Neutrino Signal Process}\label{sec:signalProc}

\begin{figure}[!t]
\includegraphics[width=0.45\textwidth]{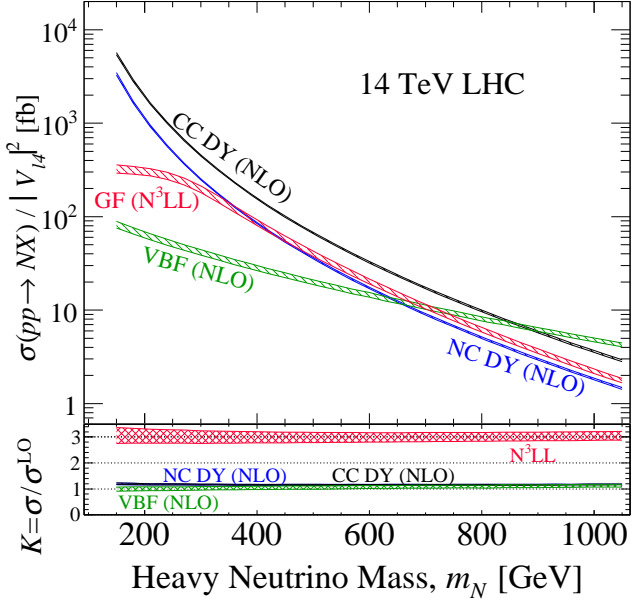}	
\caption{Upper: Leading heavy neutrino $(N)$ production mechanisms at various accuracies,
with their residual QCD scale uncertainties (band thickness),
divided by the active-sterile neutrino mixing quantity $\vert V_{\ell 4}\vert^2$,
as a function of mass $(m_N)$ at the $\sqrt{s}=14\TeV$ LHC.
Lower: QCD $K$-factor.}
\label{fig:safeVeto_XSec_vs_mN_LHCX14}
\end{figure}

In $pp$ collisions, heavy neutrinos can be produced through a variety of mechanisms that exhibit a nontrivial dependence 
on the heavy $N$ mass $m_N$ and collider energy $\sqrt{s}$.
The leading~\cite{Ruiz:2017yyf} processes include:
charged current (CC) Drell-Yan (DY), $q\overline{q}'\to N\ell^\pm$; 
neutral current (NC) DY, $q\overline{q}\to N\nu_\ell$;
$W\gamma$ fusion (VBF), $q\gamma \to N\ell^\pm q'$;
and gluon fusion (GF), $gg\to  N\nu_\ell$.
Following the procedures of Refs.~\cite{Degrande:2016aje,Ruiz:2017yyf}, 
we plot in the upper panel of Fig.~\ref{fig:safeVeto_XSec_vs_mN_LHCX14}
the production cross sections of these mechanisms at various accuracies, 
with their residual QCD scale uncertainty (band thickness) and  
divided by the mixing quantity $\vert V_{\ell 4}\vert^2$, 
as a function of $m_N$ at the $\sqrt{s}=14\TeV$ LHC.
In the lower panel is the QCD $K$-factor, defined with respect to the lowest order cross section: $K = \sigma / \sigma^{\rm LO}$.

For $m_N \sim 150\GeV-1\TeV$, we see that the sum of the CC DY and VBF production rates span about $6\pb-5\fb$ before mixing,
and translate to $6\fb-5\ab$ after taking $\left|V_{\ell 4}\right|^2 \approx 10^{-3}$, in agreement with the global fit constraints.
In light of the $\mathcal{L}=3-5\invab$ of data that will be collected over the LHC's lifetime,
including its high-luminosity phase,
the rates indicate considerable sensitivity to the mass range under discussion.
However, to date, one of the main experimental factors limiting sensitivity of multi-lepton searches for heavy neutrinos
(aside from the obvious potential to not exist) 
is the SM background associated with jets misidentified as electrons or tau leptons as well as charged leptons from non-prompt sources
~\cite{Keung:1983uu,Atre:2009rg,delAguila:2007qnc,Arganda:2015ija,Sirunyan:2018mtv}.
In what follows, we report how a jet veto, and specifically one where the $\pTVeto$ threshold is set on an event-by-event basis, can alleviate this issue.

In particular, we investigate the inclusive production of a heavy neutrino and charged lepton via the CC DY and VBF production modes,
with the subsequent decay of $N$ to only leptons, i.e.,
\begin{equation}
 p p \to \ell_N N +X \to \ell_N \ell_W W +X \to \ell_N \ell_W \ell_\nu \nu +X.
 \label{eq:pp3lX}
\end{equation}
As stipulated above, the subscript denotes the particle produced in association with the charged lepton $\ell_X$.
As a benchmark hypothesis, we assume the flavor mixing scenario,
\begin{equation}
 \vert V_{e 4} \vert = \vert V_{\tau 4} \vert \neq 0 \quad\text{and}\quad  \vert V_{\mu 4} \vert = 0,
\end{equation}
and choose the following collider signatures:
\begin{eqnarray}
 \text{Signal~I}    &:& pp \to \tau_h^\pm e^\mp \ell_X + \met,			                            \label{eq:ppTaElX} \\
 \text{Signal~II}   &:& pp \to \tau_h^+ \tau_h^- \ell_X + \met, ~\ell_X \in\{e,\mu,\tau_h\}.        \label{eq:ppTaTaX} 
\end{eqnarray}
Here, $\tau_h$ denotes a hadronically decaying $\tau$.
Due to leptonic $\tau$ decays, charged lepton flavor violation (cLFV)
cannot be unambiguously established by the simple observation of both signal processes alone. 
On the other hand, the branching rates of the $W$ and $\tau$ are known precisely.
Therefore, it is possible to falsify the no-cLFV hypothesis, 
thus deducing that $N$ couples to both electrons and $\tau$ leptons,
by accounting for how many $\tau_h^\pm e^\mp \ell_X$ events one predicts given an observed $\tau_h^+ \tau_h^- \ell_X$ rate.
Notably, the $\tau_h^\pm e^\mp \ell_X$ rate in the flavor-violating case is relatively enhanced compared to the no-flavor-violating case.

We now turn to how jet vetoes impact the signal processes in Eqs.~(\ref{eq:ppTaElX})-(\ref{eq:ppTaTaX}) and their leading SM backgrounds.

\section{Safe Jet Vetoes}\label{sec:jetVeto}

\begin{figure*}[!t]
\subfigure{\includegraphics[width=0.48\textwidth]{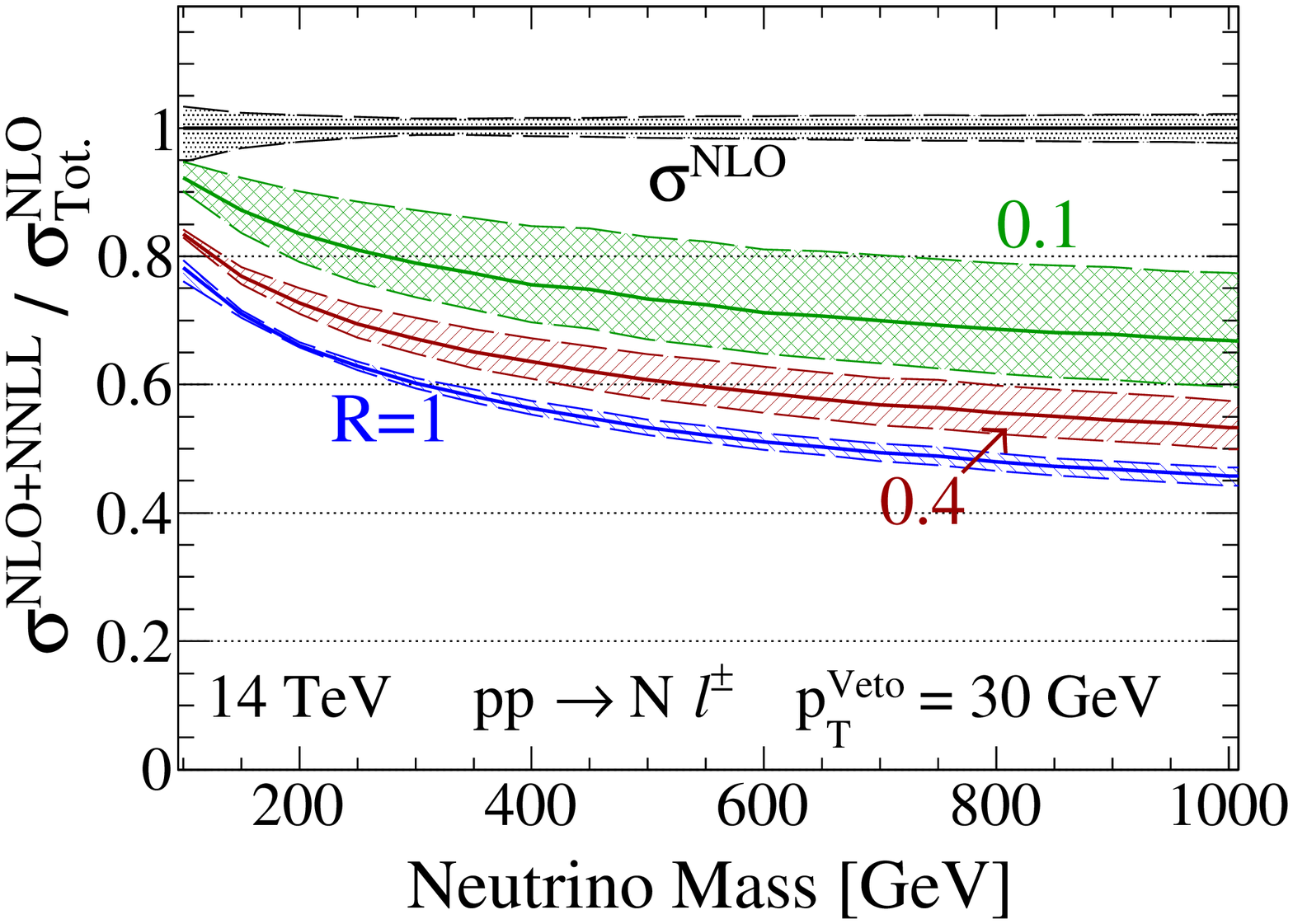}	\label{fig:vetoUnc_NLONNLL_14TeV_Static}	}
\subfigure{\includegraphics[width=0.48\textwidth]{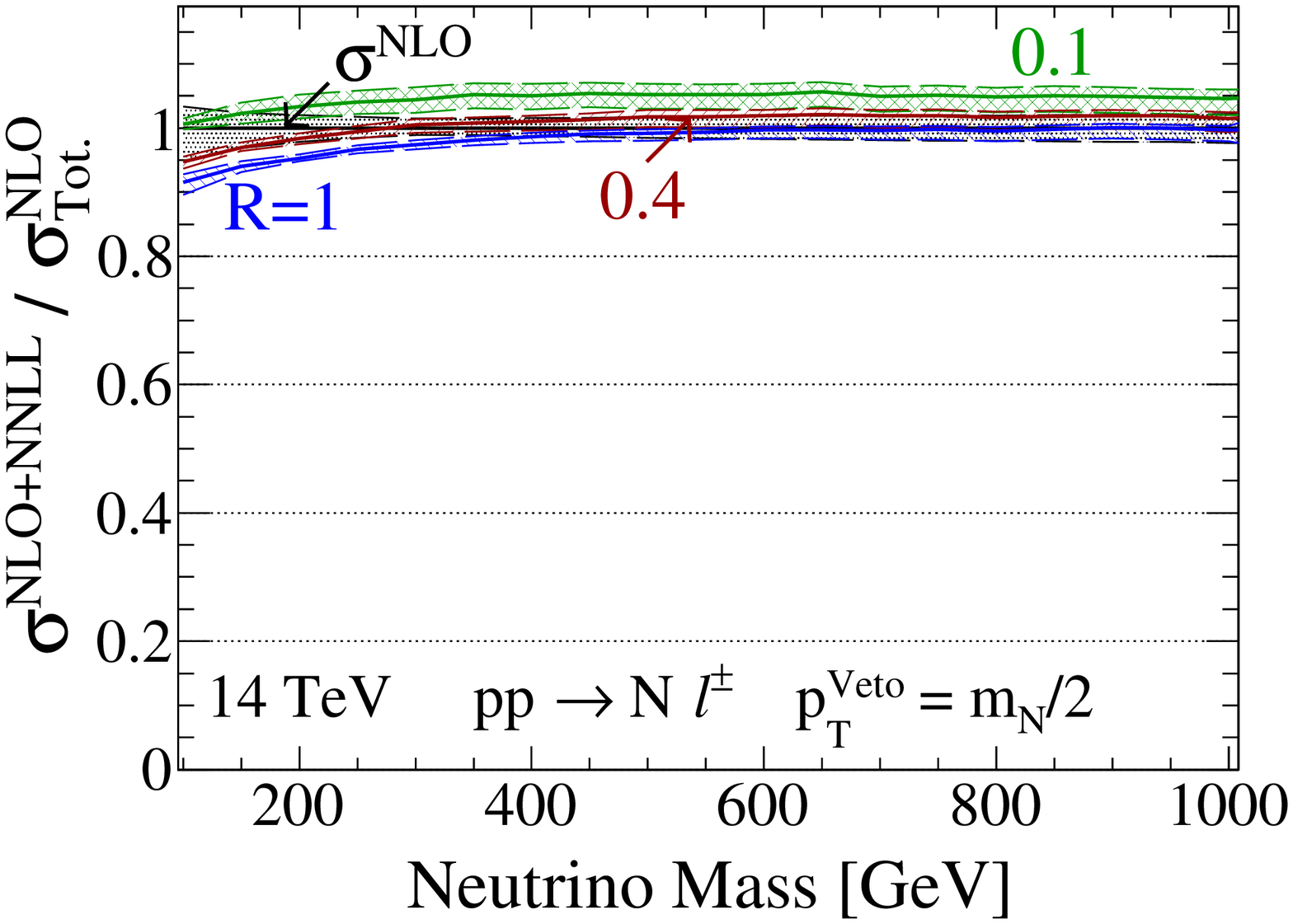}	\label{fig:vetoUnc_NLONNLL_14TeV_Dynamic}	}
\caption{Jet veto efficiencies $\varVeto$ for the $pp\to N\ell$ process as a function of heavy neutrino mass $m_N$ assuming 
(a) a static jet veto of $\pTVeto=30\GeV$ and (b) a dynamic jet veto of $\pTVeto=m_N/2$,
at $\sqrt{s}=14\TeV$ for representative jet radii $R$. The shaded areas correspond to the scale uncertainty.
}
\label{fig:vetoEffLHC14}
\end{figure*}

Central jet vetoes are premised~\cite{Barger:1990py,Barger:1991ar,Fletcher:1993ij,Barger:1994zq} on the observation that color-singlet processes,
such as Drell-Yan and EW vector boson scattering, possess characteristically different QCD radiation patterns than hard QCD processes themselves.
Color-singlet processes give rise to jets that are predominantly forward (high $\eta$) and soft (low $p_T$)
compared to those from hard QCD processes, which  are central (low $\eta$) and hard (high $p_T$).

In SM measurements, vetoes on central jets with $p_T^j>\pTVeto=25-40\GeV$ are known to yield relatively high selection efficiencies,
with the efficiencies reaching, as for example in SM $Z$ or $W^+W^-$ production~\cite{Monni:2014zra},
\begin{equation}
 \varepsilon(\pTVeto) = \sigma(p_T^j < \pTVeto)/\sigma_{\rm Tot.} \gtrsim 75-90\%.
\end{equation}
Here, $\sigma(p_T^j < \pTVeto)$ is the cross section of a signal process after the veto is applied, 
and $\sigma_{\rm Tot.}$ is the total cross section of the process before the veto.
For searches of high-mass, colorless BSM particles, however, efficiencies are known~\cite{Tackmann:2016jyb,Fuks:2017vtl} to be much lower
due to the higher likelihood to radiate high-$p_T$ gluons compared to processes with lower mass objects.
This is visible in Fig.~\ref{fig:vetoUnc_NLONNLL_14TeV_Static} where we plot
the predicted $\varVeto$ as a function of heavy neutrino mass $m_N$
for the DY process $pp\to N \ell_N$, along with their total scale uncertainties (band thickness).
We evaluate the veto at NLO+NNLL(veto) with $\pTVeto=30\GeV$ for representative jet radii $R$ and the total cross section at NLO.
One sees that $\varepsilon(\pTVeto=30\GeV)$ drops below the $80\%$ efficiency threshold 
for $R=0.1~(0.4)~[1]$ at $m_N \gtrsim 700~(150)~[100]\GeV$, with scale uncertainties spanning (roughly) $\pm10~(5)~[2]\%$.
The systematically higher veto efficiencies for smaller $R$ jets is due to 
the increasing likelihood of a jet of radius $R_0$ to split into two jets separated by a distance $R_1>2R_0$, as $R_0\to0$.
Hence, smaller radius jets are more susceptible to losing momentum through ``out-of-cone'' radiation, which increases the likelihood of surviving a jet veto.
This holds for both signal and background processes; 
for more details, see Refs.~\cite{Becher:2013xia,Tackmann:2016jyb,Fuks:2017vtl,Dasgupta:2014yra,Banfi:2015pju,Dasgupta:2016bnd} and references therein.

Despite the relatively high precision of the NLO+NNLL calculation for our Drell-Yan-type process, 
we observe that the uncertainties are acutely sensitive to the choice of jet radius.
The observed decrease in perturbative uncertainty for increasing $R$ 
is due the more inclusive nature of such jets~\cite{Dasgupta:2007wa,Becher:2013xia,Dasgupta:2014yra,Banfi:2015pju,Dasgupta:2016bnd}.
Altogether, vetoes with static choices of $\pTVeto$ result in discouraging efficiencies and uncertainties for otherwise sensible values of $\pTVeto$.

Interestingly, were one to consider the characteristic $p_T$ scales of the charged leptons in the process in Eq.~(\ref{eq:pp3lX})
one would find that each charged lepton $p_T$ scales with the mass of $N$. 
Namely, that~\cite{PRW18:toAppear}
\begin{eqnarray}
p_T^{\ell_N} 	&\sim& \frac{m_N}{3}, 	\\
p_T^{\ell_W} 	&\sim& \frac{m_N}{2} (1-M_W^2/m_N^2) \sim \frac{m_N}{2}, 				\\
p_T^{\ell_\nu} 	&\sim& \frac{m_N}{4} (1+M_W^2/m_N^2) \sim \frac{m_N}{4},	
\end{eqnarray}
where the rightmost approximations are in the $(M_W/m_N)^2\to0$ limit.
Hence, setting $\pTVeto$ to the leading, subleading, or trailing charged lepton $p_T$ (or potentially MET) does two things: 
(i) It foremost guarantees that the Sudakov logarithms $\alpha_s(\pTVeto)\log(m_N^2/\pTVeto~{}^2)$ are much less than $1$ on an event-by-event basis,
thereby reducing the need for resummation beyond LL or NLL precision.
(ii) It raises the veto threshold with increasing $m_N$, thereby countering the drop in signal efficiency due to higher jet activity.

In Fig.~\ref{fig:vetoUnc_NLONNLL_14TeV_Dynamic} we show again the veto efficiency for the $pp\to N\ell_N$ DY process but take instead $\pTVeto=m_N/2$.
(We chose this scale as a proxy for the leading charged lepton $p_T$ in order to employ the resummation formalism of Refs.~\cite{Alwall:2014hca,Becher:2014aya}.)
Remarkably, efficiencies jump to $\varVeto>90-95\%$ over the $m_N$ range considered,
with uncertainties reducing to the few percent level and exhibiting a much smaller dependence on $R$.
When $\pTVeto=m_N/4$ and $R=1$, we have checked that efficiencies remain high, 
spanning $\varVeto>90-95\%$ for $m_N\gtrsim200\GeV$ but drop to $\varVeto\sim80-85\%$ for $m_N = 150-200\GeV$.
As one may anticipate, this is comparable to the static veto since for such masses $\pTVeto = m_N/4 \sim 38 - 50\GeV$.

Interestingly, for $R=0.1$, we observe that the NLO+NNLL(veto) prediction exceeds the inclusive NLO rate, with $\varVeto = 100 - 105\%$.
The origin of this ``excess'' is an  $\mathcal{O}(\alpha_s^2)$ term in the NNLL resummation (see Ref.~\cite{Becher:2014aya} and references therein), 
and is of the form $\alpha_s^2 \log R$.
For $m_N\gtrsim150\GeV$ and jet radius $R=0.01$, we confirm that the problem worsens, with $\varVeto = 100 - 115\%$, and indicates a breakdown of the perturbative calculation.
Intuitively, the problem stems from an energetic parton in a jet of a tiny radius $R_0\ll1$ that splits into two partons separated by a still small distance $R_1 > 2R_0$
such that $2R_0 < R_1 \ll 1$. In this limit, the parton splitting is in collinear regime and should be resummed as done, for example, 
in Refs.~\cite{Dasgupta:2014yra,Banfi:2015pju,Dasgupta:2016bnd}.
This does not suggest a breakdown of our proposed dynamical jet veto scheme.
It indicates only that to quantify the impact of the veto scheme for $R\lesssim 0.2$ one needs to also resum jet radius logarithms.
While such computations are beyond the scope of the present work, it is nonetheless encouraging that with the dynamical jet veto
the dependence on the factorization and renormalization scales is now subdominant to other sources of theoretical uncertainty.

Regarding the impact of the proposed jet veto on the VBF process, which possesses at least one energetic (forward) jet with $p_T^{j_{\rm VBF}} \gtrsim M_W/2\sim40\GeV$,
it is important to emphasize that jet vetoes do not eliminate all jet activity in an event.
They remove only the high-$p_T$ jet activity.
Jet vetoes are inclusive with respect to jet activity below the $p_T^{\rm Veto}$ threshold.
This means that so long as $p_T^{j_{\rm VBF}} < p_T^{\rm Veto}$ holds, then VBF events will survive the veto.
As seen in Fig.~\ref{fig:safeVeto_XSec_vs_mN_LHCX14}, 
VBF production of heavy $N$ becomes an important channel for $m_N\gtrsim500\GeV$ 
and surpasses the DY mechanism outright for $m_N\gtrsim1\TeV$~\cite{Alva:2014gxa,Degrande:2016aje}.
Subsequently, for $m_N\gtrsim500\GeV$, one recognizes immediately that under our proposal the veto threshold scales as  $p_T^{\rm Veto}\sim 160-250\GeV$
depending on which charged lepton is actually tagged,
and surpasses the characteristic value of $p_T^{j_{\rm VBF}}$.

It is now necessary to address whether it is justifiable to exclude hadronically decaying $\tau$ leptons from the jet veto.
Experimentally, $\tau_h$ are reconstructed first as jets before $\tau$-tagging/classification~\cite{Khachatryan:2015dfa,Aad:2015unr}.
Theoretically, at some intermediate point $\tau$ leptons decay to quarks in the full, unapproximated trilepton process.
Arguably, such partons may be color-connected to the rest of the hadronic system or interfere with initial-state radiation.
Formally though, for the DY and VBF processes, such contributions appear first at $\mathcal{O}(\alpha_s^2)$, 
and hence are beyond the claimed accuracy of our calculations.
In spite of that, we note that for resonant heavy $N$ the $\tau$'s effective lifetime (ignoring abuse of notation) 
is $\tau_{\tau}\gamma_\tau \sim (1/\Gamma_\tau)(E_\tau /m_\tau) \sim m_N / (\Gamma_\tau m_\tau)$, 
where $\Gamma_\tau\sim2\times10^{-12}\GeV$.
This is much longer than the time scale of the hard process, $\tau_{\rm hard} \sim 1/m_N$.
Hence, under the narrow width approximation (NWA), which color-disconnects  the $\tau$ lepton to all orders in $\alpha_s$,
one neglects contributions of the size $\tau_{\rm hard} /(\tau_\tau\gamma_\tau) \sim (\Gamma_\tau m_\tau / m_N^2)\ll1$.
In the absence of the NWA for non-resonant $N$, however, the $\tau$'s effective lifetime may only be $\tau_{\tau}\gamma_\tau \sim 1/\Gamma_\tau$.
This too is much longer than the hadronization/non-perturbative scale, 
which is $\tau_{\rm NP} \sim 1/\Lambda_{\rm NP}$  with $\Lambda_{\rm NP} \sim 1-2$ GeV.
Hence, the $\tau$ lepton outlives the primary hadronization and exchanges between $\tau_h$ and the remainder of the hadronic system are
long range color-singlet exchanges~\cite{Sjostrand:1993rb,Khoze:1994fu,Khoze:1998gi,Collins:2011zzd,Gaunt:2014ska}, 
i.e., higher twist, and hence beyond the accuracy of the Factorization Theorem itself.

Legitimately, one may question if such a veto also dramatically and incidentally increases the acceptance rates of QCD backgrounds.
As we now discuss, it does not.

\begin{figure*}[!t]
\subfigure{\includegraphics[width=0.48\textwidth]{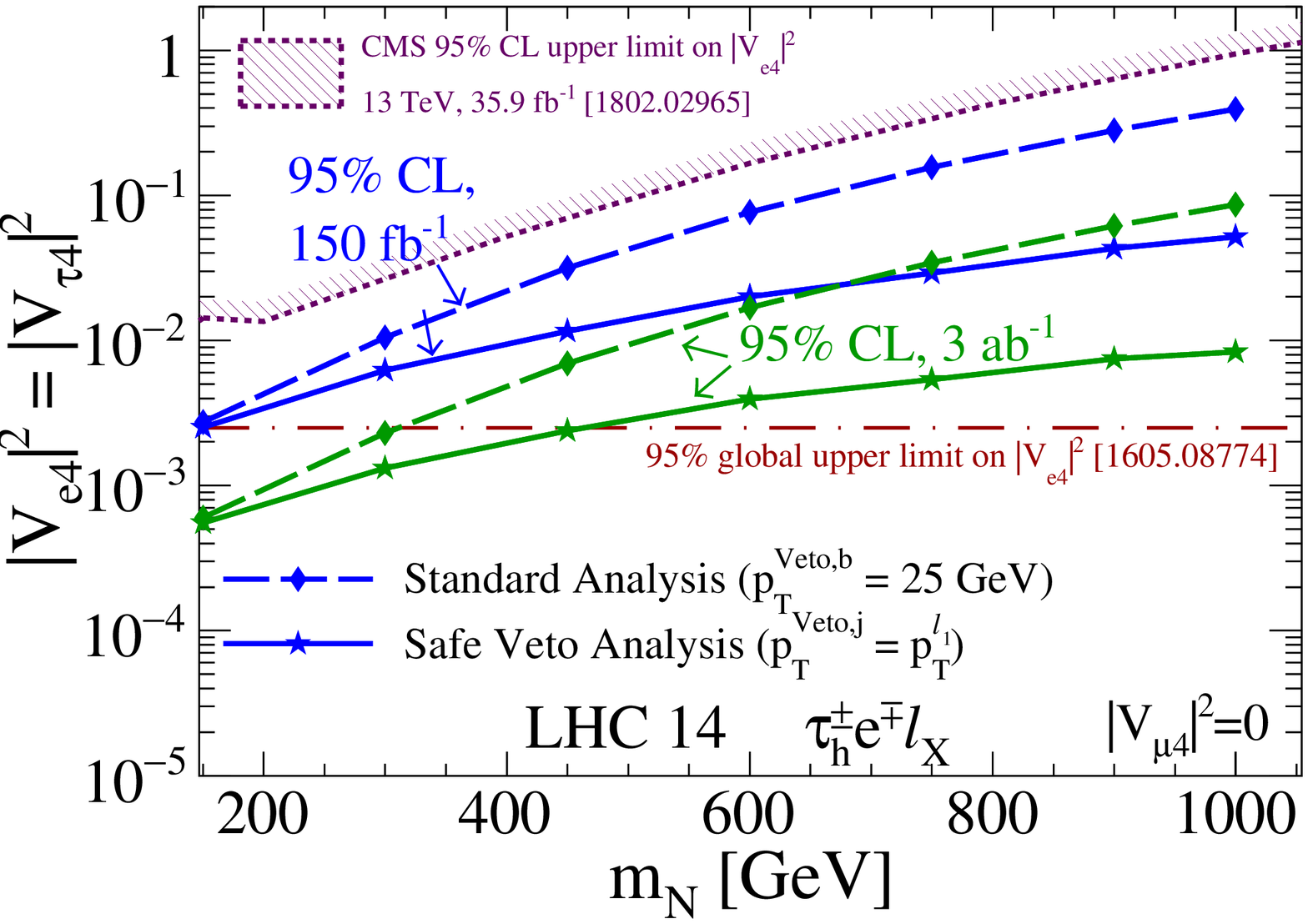}	\label{fig:sElElVsMN14TeV}	}
\subfigure{\includegraphics[width=0.48\textwidth]{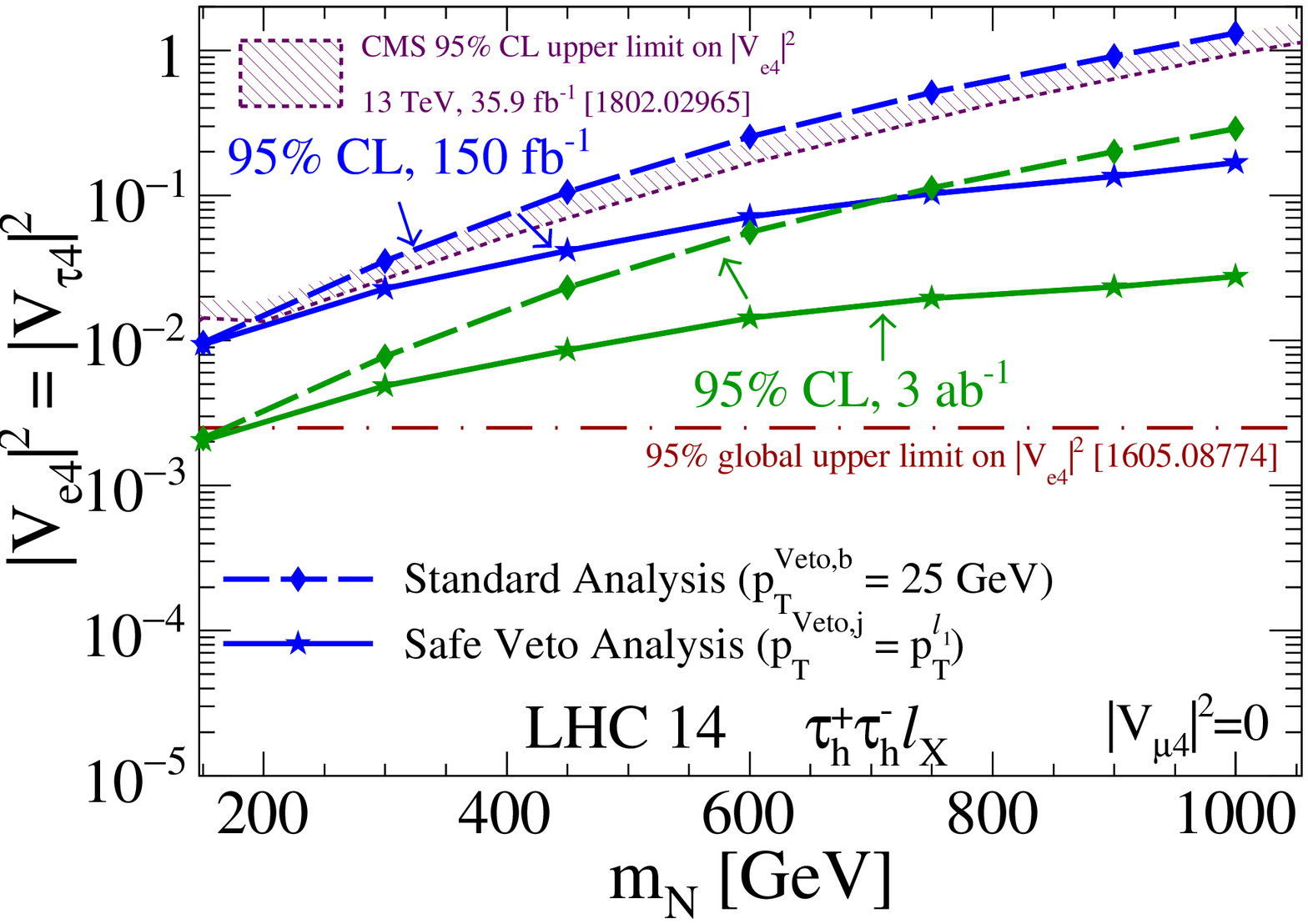}	\label{fig:sMuMuVsMN14TeV}	}
\caption{
The 14 TeV LHC 95\% CL sensitivity to heavy Dirac neutrino $(N)$ active-sterile mixing,
assuming $\vert V_{e4}\vert^2 = \vert V_{\tau4}\vert^2$ and $\vert V_{\mu4}\vert^2=0$,
as a function of $N$ mass $(m_N)$ [GeV],
via the trilepton final states (a) $\tau_h^\pm e^\mp X_\ell$ and (b) $\tau_h^+\tau_h^- X_\ell$, for $X_\ell \in\{e,\mu,\tau\}$,
using the standard analysis with a b-jet veto (dash-diamond) and proposed jet veto-based analysis (solid-star)
after $\mathcal{L}=150\invfb$ and $3\invab$.
Also shown are the 95\% CL limits on $\vert V_{e4}\vert^2$ from CMS at 13 TeV~\cite{Sirunyan:2018mtv} and global constraints~\cite{Fernandez-Martinez:2016lgt}.
}
\label{fig:mixingSensitivity}
\end{figure*}

\subsubsection*{Top Quark Production}
Due to their inherent mass scales and rates, single and pair production of top quarks are major backgrounds to any measurement and search 
for EW and colorless BSM processes in TeV-scale hadron collisions.
However, as investigated in Ref.~\cite{Fuks:2017vtl}, the $p_T$ distribution of the leading jet for top quark and Drell-Yan processes 
are qualitatively different, even after the application of a veto on $b$-jets.
This implies that flavor-inclusive vetoes can generically~\cite{Fuks:2017vtl} improve signal-to-background ratios over flavor-exclusive vetoes,
a conclusion that also holds for the types of vetoes considered here:
While the characteristic $p_T$ of a charged lepton in the $t\to Wb \to \ell\nu_\ell b$ transition 
scales as $p_T^\ell \sim E_W/2 = m_t(1+M_W^2/m_t^2)/4 \approx 50-55\GeV$, 
the $b$'s $p_T$ scale is larger with $p_T^b\sim m_t(1-M_W^2/m_t^2)/2 \approx 65-70\GeV$.
The issue is more extreme for $t(\overline{t})V,$ with $V\in\{W,Z\}$, where sub-leading and trailing leptons 
possess momenta that scale as $p_T^\ell \sim M_V/2 \sim 40-45\GeV$, which is again lower than $p_T^b$.

\subsubsection*{EW Triboson Production}
The production of three (or more) EW bosons represents the main background that survives after traditional selection cuts
but actually is particularly vulnerable to the veto.
NLO corrections reveal~\cite{Campanario:2008yg,Binoth:2008kt} that $\mathcal{O}(30\%)$ of the inclusive $pp\to3W+X$ process is made of the 
$3W+1j$ subprocess; the remaining  is Born-like. 
Hence, the veto imposes a non-negligible selection cut and restricts the intermediate $W$s to be largely at rest
since recoiling against jets must be split six ways amongst the $W$s' decay products.
The scalar sum over the charged lepton $p_T^\ell$ therefore possesses a characteristic value of
\begin{equation}
 S_T^{3W} \equiv \sum_{\ell} \vert \vec{p_T}^\ell \vert \sim 3\frac{M_W}{2} \sim 120\GeV.
\end{equation}
For the $N$ mass range we consider $(m_N>150\GeV)$, one sees that the signal process characteristically exceeds this, with
\begin{equation}
 S_T^{N}  \sim \frac{m_N}{3} + \frac{m_N}{2} + \frac{m_N}{4} = \frac{13}{12}m_N.
\end{equation}
Therefore, a jet veto in conjunction with $S_T>120\GeV$ will suppress such background processes.

\subsubsection*{EW Diboson Production}
Resonant EW diboson production can be stymied by standard invariant mass cuts,
\begin{eqnarray}
 m_{\ell_i\ell_j}>10\GeV, & &  \vert m_{\ell_i\ell_j} - M_Z \vert > 15\GeV, \nonumber\\
 \text{and} & &  \vert m_{3\ell} - M_Z \vert > 15\GeV,
 \label{cut:lepInvMass}
\end{eqnarray}
on any combination of analysis-level charged leptons.
Indiscriminate application to all $\ell_i\ell_j$ helps suppress charge mis-measurement and fake lepton backgrounds.
Highly non-negligible, non-resonant contribution to the inclusive $pp\to \ell^+\ell^-\ell^\pm\nu$ and $\ell^+\ell^-\ell^+\ell^-$ processes
can be sufficiently reined in by the veto+$S_T$ selections.

\subsubsection*{Fake Leptons}
Non-prompt leptons from heavy quark decays, light jets mis-tagged as hadronic decays of $\tau$ leptons,
and light jets misidentified as electrons, collectively labeled as ``fake leptons,'' 
represent the second most important background in searches for heavy neutrinos at the LHC~\cite{Sirunyan:2018mtv}.
In most instances of fake leptons, however, a degree of high-$p_T$ QCD activity is required.
Invariably, the presence of a central, energetic jet implies, by color conservation,
that its progenitor parton is color-connected to some other part of the collision.
Hence, whether the additional colored particles constitute the beam remnant or the hard process,
there is a high likelihood that the fake lepton is accompanied by a real jet of comparable $p_T$.
This is especially the case for semi-leptonic decays of heavy flavor hadrons, e.g., $\mathcal{B}\to\mathcal{D}\ell\nu$,
where the hadronic and leptonic decay products carry comparable momenta~\cite{Isgur:1988gb,Scora:1995ty}.

In closing this section, we summarize and reiterate that implementing a dynamical jet veto scheme in searches for heavy neutrinos decaying to a multi-lepton final state 
exhibits rather desirable theoretical and phenomenological properties.
The arguments invoked rely on color-flow and a careful assessment of scales involved in the hard scattering processes (DY and VBF) 
as well as backgrounds (including ``fake'' leptons), and hence should be applicable to other high-mass, colorless processes.
We encourage investigations into applicability in the context of jet radii $R\ll1$, as well as analogous dynamical veto schemes, 
such as ones based on momentum imbalance or leading photon $p_T$.
In light of finding greatly reduced theory uncertainties as well as potential for improving signal survival and background rejection efficiencies,
we refer to this veto scheme as a safe jet veto.

Before reporting our results, we briefly comment on the ability to implement the proposed jet veto scheme in experimental searches for new, colorless particles at hadron colliders.
In particular, we anticipate that complications surrounding the delicate procedures of jet energy calibration and signal region validation are already surmountable.
Jets are calibrated with $Z+nj$ and prompt photon/$\gamma+nj$ samples in a largely analysis-independent manner~\cite{Khachatryan:2016kdb,Aaboud:2017jcu}, 
including jets with radius up to $R=1$~\cite{Aaboud:2018kfi}.
$Z+nj$ events are specifically excluded from our signal region using the standard charged lepton invariant mass cuts given in Eq.~(\ref{cut:lepInvMass}).
Moreover, dynamically defined signal regions have already been used in the context 
of searches for resonant production of Higgs boson pairs at $\sqrt{s}=13$ TeV~\cite{Aaboud:2018knk}.
While construction of dynamically defined signal, validation, and control regions represent nontrivial tasks for the present case, 
ongoing work, for example as in Ref.~\cite{Aaboud:2018knk}, by the ATLAS and CMS collaborations is very encouraging.

\section{Results and Outlook}\label{sec:results}

In Sec.~\ref{sec:jetVeto} we discussed the signal and background phenomenology for the heavy $N$ trilepton process under a dynamical jet veto.
We now report quantitatively how a search analysis designed and centered around a dynamical jet veto can improve LHC sensitivity compared to a
more traditional analysis premised solely on the presence of high-$p_T$ charged leptons and vetoes only $b$-tagged jets, 
e.g., Refs~\cite{delAguila:2007qnc,Sirunyan:2018mtv}.
We observe that simply adding a dynamical veto to a traditional analysis does not significantly improve discovery potential.

For further details and motivation of the following selection analysis, see Ref.~\cite{PRW18:toAppear}.
We define analysis-quality charged leptons and jets as isolated objects satisfying the following fiducial and kinematic cuts:
\begin{eqnarray}
  p_T^{e~(\mu)~[\tau_h]~\{j\}} > 15~(15)~[30]~\{25\}\GeV  	~\text{with}~\qquad \nonumber\\
 \vert \eta^{\mu,\tau_h,j} \vert < 2.4, ~\text{and}~ \vert \eta^e \vert < 1.4  ~\text{or}~ 1.6 < \vert \eta^e \vert < 2.4.
 \label{cut:FidKin}
\end{eqnarray}
Charged leptons and jets are then labeled according to $p_T$, with $p_T^k > p_T^{k+1}$, and the missing transverse momentum vector $\not\!\!\vec{p}_T$
is built from all visible momenta above 1 GeV in the fiducial region.
To simulate detector effects, momentum smearing is done as in Ref.~\cite{Fuks:2017vtl};
$p_T$-based (mis)tagging, (mis)identification, and fake lepton efficiencies are based on
the Detector Performance (DP) and dedicated studies of 
Refs.~\cite{Alvarez:2016nrz,Sirunyan:2017uzs,Sirunyan:2017ezt,CMS-DP-2017-036,CMS-DP-2018-009}.
We require events to contain exactly three analysis-quality charged leptons with flavor composition according to Eqs.~(\ref{eq:ppTaElX})-(\ref{eq:ppTaTaX}).
We next apply the invariant mass cuts of Eq.~(\ref{cut:lepInvMass}).
After imposing a jet veto set to the $p_T$ of the leading charged lepton, i.e., $\pTVeto=p_T^{\ell_1}$, we impose that $S_T > 120\GeV$.
As a proxy to the invariant mass of $N$, we build a version of the multi-body transverse mass~\cite{Barger:1983wf,Barger:1988mr},
\begin{eqnarray}
\tilde{M}_{T,i}^2 &=& \left[\sqrt{p_T^2(\ell^{\rm OS}) + m_{\ell^{\rm OS}}^2} + \sqrt{p_T^2(\ell_i^{\rm SS},\not\!\vec{p}_T) + M_W^2}\right]^2
\nonumber\\
&-& \left[\vec{p}_{T}(\ell^{\rm OS},\ell_i^{\rm SS}) + \not\!\vec{p}_T \right]^2, \quad i=1,2.
\end{eqnarray}
Here, $\ell^{\rm OS}~(\ell^{\rm SS}_i)$ is the one opposite-sign (either same-sign) charged lepton in the trilepton final-state.
Of the two permutations of $\tilde{M}_{T,i}$, we choose the one $(\hat{M}_{T})$ closest to our mass hypothesis  
and select for events satisfying
\begin{equation}
 -0.15 < \frac{(\hat{M}_{T} - m_N^{\rm hypothesis})}{m_N^{\rm hypothesis}} < 0.1.
 \label{cut:NuMass}
\end{equation}

As a benchmark, we base the ``standard analysis'' on the 13 TeV CMS search for heavy neutrinos~\cite{Sirunyan:2018mtv}.
Starting from Eqs.~(\ref{cut:FidKin}) and ~(\ref{cut:lepInvMass}), 
and assuming the same flavor combinations as before, we require that
\begin{eqnarray}
p_T^{\ell_1} > 55\GeV, ~p_T^{\ell_2}>15\GeV, ~m_{3\ell}>80\GeV.
\end{eqnarray}
Events with at least one $b$-tagged jet are vetoed. 
The results of our traditional analysis are in line with Ref.~\cite{Sirunyan:2018mtv}.

Assuming Gaussian statistics and a background systematic weighting of $B\to(1+\delta_B B)$, with $\delta_B=10\%$,
we show in Fig.~\ref{fig:mixingSensitivity} the 95\% CL sensitivity to the active-sterile mixing quantity $\vert V_{e 4}\vert^2,~\vert V_{\tau 4}\vert^2$ 
in the (a) $ \tau_h^\pm e^\mp \ell_X$ and (b) $\tau_h^+ \tau_h^- \ell_X$ final state,
for the veto (solid-star) and standard (dash-diamond) analyses, at the 14 TeV LHC with $\mathcal{L}=150\invfb$ and $3\invab$ of data.
The improvement in sensitivity when applying the jet veto is unambiguous. 
We find that the veto can increase the reach of $\vert V_{\ell 4}\vert^2$ by 
up to a factor of 7-8 with $150\invfb$ and up to a factor of 10-11 with $3\invab$. 
Hence, with $3\invab$, LHC searches can surpass indirect limits on the active-heavy mixing obtained from global fits to EW precision observables and low-energy data.
In summary, with $\mathcal{L}= 150~\invfb~(3\invab)$, we find $\vert V_{e4}\vert^2, \vert V_{\tau 4}\vert^2 < 2.5\times10^{-3}-5.2\times10^{-2}~(5.5\times10^{-4}-8.3\times10^{-3})$ 
can be probed at the 95\% for $m_N=150-1000$ GeV at the 14 TeV LHC.
We stress that the improvement at high $m_N$ stems both from an increase in signal rate and a decrease in background rate.
At low $m_N$, however, small improvement is observed, in part, due to the stringent $p_T$ requirements for $\tau_h$ tagging,
which depletes signal strength despite improved efficiencies.

Reporting the impact on other flavor combinations is beyond our present scope and refer readers to Ref.~\cite{PRW18:toAppear}.

\section{Summary and Conclusion}\label{sec:conclusions}

Due to inherently different radiation patterns, jet vetoes are powerful techniques to reduce QCD backgrounds in measurements and searches 
for electroweak and color-singlet new physics processes in hadron collisions. 
In this letter, we report key findings when premising a search strategy on
vetoing events with jets possessing transverse momenta $(p_T)$ greater than the highest $p_T$ charged lepton in the event.
We demonstrate that they can be successfully used in a broad class of experimental searches,
including searches for new high-mass particles as well as events with $\tau$ leptons decaying hadronically.
We find an impressive improvement over traditional, fixed-$p_T$ jet vetoes.

As a representative case study, we focused on the impact of jet vetoes in searches for heavy Dirac neutrinos $(N)$
participating in the trilepton process $pp\to \ell_N N \to \ell_N \ell_W W \to \ell_N \ell_W \ell_\nu \nu$.
The phenomenological consequences of such a jet veto on the signal and background processes are summarized in Sec.~\ref{sec:jetVeto}.
We find the following: 
(i) As shown in Fig.~\ref{fig:vetoEffLHC14}, perturbative uncertainties associated with the veto greatly reduce 
due to tying the veto scale to the hard process scale.
(ii) Also shown in the figure is that the signal efficiency for passing the veto exceeds $90-95\%$ 
for $N$ with masses in the range $m_N=150-1000$ GeV, and exhibits little-to-no dependence on the neutrino mass scale.
(iii)  Top quark and ``fake'' lepton rejection capabilities also improve compared to only vetoing heavy flavor-tagged jets.
Subsequently, as shown in Fig.~\ref{fig:mixingSensitivity}, this results in an improved sensitivity to the heavy neutrino mixing
quantity $\vert V_{\ell 4}\vert^2$ up to an order of magnitude over the LHC's lifetime; see Sec.~\ref{sec:results}.
For a Dirac neutrino with mass $m_N = 150-1000$ GeV and the representative active-sterile mixing hypothesis 
$\vert V_{e4}\vert = \vert V_{\tau 4}\vert$ with $\vert V_{\mu 4}\vert=0$,
we report that LHC experiments can probe $\vert V_{e4}\vert^2, \vert V_{\tau 4}\vert^2 \lesssim 6\times10^{-4} - 8\times10^{-3}$.
Further investigations into the impact on heavy neutrino searches in different flavor channels is left to future work~\cite{PRW18:toAppear}.
We anticipate that sensitivity could be further improved if combined with advanced multivariate techniques, and encourage future work on the topic.

\section{Acknowledgements}

Julien Baglio, Thomas Becher, Agni Bethani, Lydia Brenner, Tom Cornelis,
Mrinal Dasgupta, Didar Dobur, Tao Han, Josh Isaacson, Valery Khoze, Valerie Lang, Kate Pachal,
Kevin Pedro, Stefan Prestel, Zhuoni Qian, Jakub Scholtz, Lesya Shchutska, Martin Fibonacci Tamarit, and Xing Wang are thanked for discussions.
This work was funded in part by the UK STFC, and the European Union's Horizon 2020 research and innovation programme 
under the Marie Sklodowska-Curie grant agreements No 690575 (RISE InvisiblesPlus) and No 674896 (ITN ELUSIVE).
SP and CW receive financial support from the European Research Council under the European Unions Seventh Framework
Programme (FP/2007-2013)/ERC Grant NuMass Agreement No. 617143.
SP acknowledges partial support from the Wolfson Foundation and the Royal Society.
RR and CW acknowledge the hospitality of the University of Pittsburgh's PITT-PACC during the completion of this work.
RR acknowledges the hospitality of Fermilab and DESY.


\end{document}